\documentstyle{article}

\begin{document}
\centerline{QUANTUM MEASUREMENT PROBLEM} 
\centerline{AND THE POSSIBLE 
ROLE OF THE GRAVITATIONAL FIELD}
\vskip 2cm
\centerline{J. Anandan}
\centerline{Institute for Advanced Studies}
\centerline{Hebrew University of 
Jerusalem}
\centerline{Givat Ram, Israel 91904}
\centerline{and}
\centerline{Department of Physics and Astronomy}
\centerline{University of 
South
Carolina}
\centerline{Columbia, SC 29208, USA.}
\centerline{E-mail: jeeva@sc.edu}

\begin{abstract}
The quantum measurement problem and various unsuccessful 
attempts to resolve 
it are reviewed. A 
suggestion by Diosi and Penrose for the half 
life of the quantum superposition of two Newtonian gravitational 
fields is generalized to an arbitrary quantum superposition 
of relativistic, but weak, gravitational fields. The nature of the 
``collapse'' process of the wave function is examined.
\end{abstract}

\section{Introduction}

Two of the most important unsolved problems in theoretical 
physics are the problem of quantizing gravitation and the 
measurement problem in quantum theory. It is possible that the 
solution of each one needs the other. Since we have successful 
quantum theories of electroweak, and strong interactions, the 
solution to the problem of the collapse of the wave function, 
known 
as the measurement problem, may lie in the yet unknown 
quantum 
theory of the remaining interaction, namely quantum gravity. On 
the 
other hand, the numerous unsuccessful attempts to construct a 
quantum theory of gravity for more than six decades on the 
assumption that the present linear quantum theory is correct 
suggests that perhaps not only general relativity but also quantum 
theory should be modified in order to construct of a satisfactory 
quantum theory of gravity.

In section 2, I shall briefly review the measurement problem and 
protective 
observations.  I shall argue, in section 3, that 
none of the standard interpretations of quantum theory provide a 
solution for the measurement problem. This suggests that a 
modification of 
quantum theory 
may be needed, particularly since protective observation suggests 
that the 
wave function is real and therefore the reduction of the wave 
packet 
during 
measurement is a real objective process. I then consider the 
specific 
suggestion along the 
lines mentioned above due to Roger Penrose 
\cite{pe81,pe86,pe89,pe93,pe96}. He advocated the use of
the gravitational field of the wave function to explain its reduction 
during measurement. Several other physicists have also 
argued that the phenomenon of state vector reduction is an 
objective, 
real process, and not just a change in the state of knowledge of the 
observer \cite{ka66,pea86,grw,pea89,ggr90}, an important aspect of 
Penrose's 
proposal and that of Diosi \cite{di87,di89} is that 
they have a quantitative prediction for the time of collapse of the 
wave 
function, which has the potential of being subject to experimental 
tests. Conversely, experiments 
could 
guide us 
in 
constructing a definite theory (which still does not exist) that 
would 
justify or modify this proposal.

\section{Quantum measurement problem}

The simplest, though dramatic, statement of the 
measurement problem in quantum theory is that {\it quantum 
theory does not explain the occurrence of events}. So, quantum 
theory does not explain the first thing we observe about the world 
around us.

To see this, consider a quantum system whose state is described 
by a 
wave 
function $\psi$ just before it interacts with an apparatus, which 
we 
shall treat quantum mechanically also. Suppose $\psi = \sum_i 
c_i\psi_i$, where $\psi_i$ is a state of the system which after it 
interacts with the apparatus leaves it in the state that is described 
by the wave function $\psi'_i\alpha_i$, where $\psi'_i$ represents 
the new state of the system and $\alpha_i$ the corresponding 
state 
of the apparatus. We represent this by 
\begin{equation}
\psi_i \alpha \rightarrow \psi'_i\alpha_i 
\label{eq:spot}
\end{equation}
Then it follows from the linearity of quantum evolution that the 
interaction of $\psi$ with the screen is represented by 
\begin{equation}
\psi\alpha \rightarrow \sum_i c_i \psi'_i\alpha_i .
\label{eq:entangle}
\end{equation} 
The resulting state is called an entangled state, meaning that it 
cannot be written as simple product of the form $\psi'\alpha'$.

For example, the quantum system may be a photon and the 
apparatus a photographic plate. Then $\psi_i$ is a localized wave 
function of the photon which interacts with the plate to trigger a 
chemical reaction which results in a spot on the screen 
represented 
by $\alpha_i$. 
But $\psi$ produces a quantum superposition of many different 
spots on the screen that correspond to the different states $\psi_i$ 
of 
the photon. Since the photon has now been absorbed, 
$\psi'_i\alpha_i 
$ in the right hand side of (\ref{eq:spot}) and (\ref{eq:entangle}) 
may 
be 
replaced simply by 
$\alpha_i$. We would not call the resulting state entangled. 
Nevertheless, the experiments (\ref{eq:spot}) establishes a 
correlation 
between 
the state $\psi_i$ and $\alpha_i$. So, if 
we observe $\alpha_i$, whether or not the quantum system is 
now 
present, we can deduce the state of the system $\psi_i$ that 
would 
have 
caused this state of the apparatus. So, what is essential to 
measurement is 
this {\it 
correlation} between the system states and the corresponding 
apparatus states and not entanglement.

However, for a given photon in the state $\psi$ we actually 
observe 
only one spot described, say, by $\alpha_k$. This appearence of a 
spot may be regarded as an approximate representation of an 
event, 
because it occurs in a fairly localized region of space-time that is 
defined by the small spatial region on the screen and the small 
interval of 
time during which it is formed.  But (\ref{eq:entangle}) by itself 
does not explain the appearance of this `event'.  So, we 
need to make an additional `projection postulate'
\begin{equation} 
\sum_i c_i \psi'_i\alpha_i \rightarrow \psi'_k\alpha_k ,
\label{eq:collapse}
\end{equation}
where $\psi'_k\alpha_k$ represent the particular event or set of 
events 
observed. The quantum measurement problem is the problem of 
understanding 
(\ref{eq:collapse}), which is referred to as the reduction of the 
wave 
packet 
or 
collapse of the wave function. For example, is (\ref{eq:collapse}) 
an 
objective 
dynamical process, which we may take (\ref{eq:entangle}) to be, 
or is 
it a 
subjective process we make in our minds due to the additional 
information we obtain from the measurement? Or what 
determines 
the {\it preferred states} $\alpha_i$ into which the reduction 
takes 
place?

So, the state vector undergoes two types of changes \cite{wi63}, 
which using the terminology of Penrose \cite{pe89,pe96}, may be 
called the $U$ and $R$ processes.  
to which (\ref{eq:spot}) and (\ref{eq:entangle})  are examples of 
what he 
calls the $U$ 
process, whereas (\ref{eq:collapse}) is the $R$ process .  The $U$ 
process is 
the 
linear unitary evolution which in the present day quantum theory 
is 
governed by Schr\"odinger's equation. But what causes the 
measurement problem is the linearity of the $U$ process. The 
unitarity is really relevant to the $R$ process. Unitarity ensures 
that 
the 
sum of the probabilities of the possible outcomes in any 
measurement, each of which is given by an $R$ process remains 
constant during the $U$ time evolution. This of course follows 
from 
the postulate that the transition probability from the initial to the 
final state in the $R$ process is the square of the modulus of the 
inner product between normalized state vectors representing the 
two 
states. 

The process of measurement, as described above, takes place in 
two 
stages: First is the entanglement (\ref{eq:entangle}) and the 
second is 
the 
collapse 
(\ref{eq:collapse}). If we had no choice in preparing the initial 
state 
of the 
system 
then $\psi$ is in general a superposition of the $\psi_i$s. Then the 
entanglement (\ref{eq:entangle}) is the inevitable consequence of 
the 
linearity of 
the evolution. But if we could prepare the state then it is possible 
to 
prevent entanglement as in the case of protective 
observation \cite{aav93}. I.e. in such an observation
\begin{equation}
\psi\alpha \rightarrow \psi'\alpha' ,
\label{eq:protective}
\end{equation}
where the state represented by $\psi'$ does not differ appreciably 
from the state $\psi$. The protection is usually an external 
interaction which puts $\psi$ in an eigenstate of the Hamiltonian 
and 
the measurement process results in adiabatic evolution. 

Then $\alpha'$ gives information about $\psi$; specifically it tells 
us 
the `expectation value' with respect to $\psi$ of the obervable of 
the 
system that it is coupled to an apparatus observable. By doing 
such 
experiments a large number of times it is possible to determine 
$\psi$ (up to phase) even though the system is always undergoing 
$U$ evolution. Consequently, the statistical interpretation of 
quantum 
mechanics is avoided during protective observations. Indeed, 
$\psi$ 
may be determined using just one system which is subject to 
many 
experiments.

If the protection mechanism is precisely known then it would be 
possible to determine the $\psi$ by means of calculation. But 
there is 
a profound difference between experimentally observing the 
state, 
which gives the manifestation of the state, and calculating it. Also, 
the protected state need not be in an eigenstate of the observable 
being measured, and yet there is no entanglement. It may appear 
that if the 
combined system evolves as (\ref{eq:protective}), as in a protecive 
measurement, then we cannot obtain new information about the 
system 
state, because if this state were previously unknown then there 
should be 
the possibility of the system being in more than one state with 
respect to the 
apparatus, i.e. there should be entanglement or correlation 
between 
system 
states and apparatus states. This is true if the system states 
already 
have a 
well defined meaning. 

However, a 
state acquires meaning through its relation to other states. E.g. 
describing a state vector $|\psi>$ by means of its wave function  is 
the same as giving the inner product of $|\psi>$ with all the 
eigenstates of the position operator. A previously proved 
theorem \cite{an93} states that from the `expectation values' 
defined 
as functions on the set of physical states, it is possible to construct 
the Hilbert space whose rays are these states. Indeed the entire 
machinery of quantum mechanics may be constructed from the 
numbers which an experimentalist obtains by protective 
measurements. Before the Hilbert space is reconstructed, it is not 
possible to calculate the wave function. However, from the 
information which can in principle be obtained from protective 
measurements, it is possible to determine the inner products 
between a given state vector and all other state vectors, according 
to 
the 
above mentioned theorem, which gives 
meaning to the given state vector. So if the meanings of the states 
are 
previously unknown, then in this way it is possible to obtain 
new information that determine the state of a system by 
means of protective observations, even though the evolution is 
according to 
\ref{eq:protective}. Also, this is done by using just one quantum 
system in 
the given state. No statistical interpretation of the wave function is 
needed. 
This suggests that the wave function may be real and objective.

\section{Efforts to resolve the measurement problem}

The well known Copenhagen interpretation attempts to deal with 
the 
measurement problem by introducing an artificial division 
between 
the quantum system being observed and the apparatus. The 
quantum system, which was assumed to be `microscopic', is 
treated 
quantum mechanically. Its state evolves in a Hilbert space. The 
apparatus, assumed to be `macroscopic', is treated classically. The 
discontinuous $R$ process occurs when the microscopic system 
interacts with the macroscopic system. This is accounted for by 
supposing that the wave function represents only our knowledge 
of 
the state of the system, and this knowledge undergoes a 
discontinuous change when the measurement is made.

This is unsatisfactory because the apparatus is made up of 
electrons, 
protons, neutrons and photons, which are clearly quantum 
mechanical. At the time when the Copenhagen interpretation was 
formulated, it was not known that a macroscopic superconductor 
must be treated quantum mechanically. Also, Bose-Einstein 
condensation, which provides another clearly macroscopic 
quantum 
system, was not experimentally realized. Today we know and 
possess 
macroscopic quantum systems. Moreover, there have been 
numerous 
quantum mechanical experiments on a macroscopic scale, using 
superconductors, electron, neutron and atomic interferometry. 
Another related problem is that the Copenhagen interpretation 
does 
not specify the line of division between the system and the 
apparatus. It does not give a number, specifying the complexity or 
mass of the system, which when exceeded would make the system 
macroscopic. Also, the early universe needs to be treated quantum 
mechanically because quantum gravitational effects were so 
important at that time. But nothing can be more macroscopic than 
the universe. And the universe is everything there is, so no line of 
division can be specified between it and the apparatus.
Finally, protective observation, discussed in section 1, 
suggests that the wave function is real and objective, and is not 
just 
our 
knowledge of the system.

This brings us to two famous interpretations of quantum theory in 
which the wave function is regarded as real, consistent with the 
meaning of the wave function given by protective observation. 
One is 
the Everett interpretation \cite{ev57} in which the wave function 
never collapses. But this view carries with it a huge excess 
intellectual baggage in the form of infinitely many worlds that 
coexist with the world in which we observe ourselves in. Also, it 
does 
not seem to explain the `preferred basis' or the `interpretation 
basis' 
in which we observe the world to have a fairly classical space-
time 
description. The latter description is very different from the 
Hilbert 
space description which is the only reality in the Everett view. 
Furthermore, since the Everett view gives a deterministic 
description 
of a real state vector, the only natural way of introducing 
probabilities is by coarse graining. But this would not agree with 
the 
probabilities determined by the inner product in Hilbert space 
which 
is well confirmed by experiment. 

The Bohm interpretation~\cite{bo52}, tries to overcome the old 
problem of wave-particle duality that asserts the simultaneous 
existence of both the particle and the wave. This dual ontology 
enables one to have the cake and eat it too. A direct experimental 
evidence of a particle, such as the triggering of a particle detector, 
or 
a track in a cloud chamber, etc. is explained as caused by the 
particle. 
And this is the {\it only} role of the particle. The motion of the 
particle is assumed to be guided by a `quantum potential' which 
explains, for example, the result of an interference experiment. 
Without the particle the Bohm interpretation would be like the 
Everett interpretation in that there is no collapse of the wave 
function. But the particles determine which branch of the wave 
function we {i.e. the particles constituting us) are in. So, there is no 
excess 
intellectual baggage of the 
many worlds as far as the particles are concerned. But there are 
the 
`empty waves' of the other branches. These waves may be 
protectively observed and therefore may be regarded as 
real~\cite{an95}. This has the advantage over the previous 
interpretations in that there is no preferred basis problem 
because 
the particles determine `events', e.g. spots on the photographic 
plate, 
which give the illusion of a preferred basis in the Hilbert space.

The absence of any further role for the particle is illustrated by 
the 
fact that the particle does not react back on the wave. This 
violates 
the action-reaction principle, which may be regarded as a 
metaphysical objection to the Bohm interpretation
~\cite{ho93,an95}. It is 
also strange that in this theory the wave function plays a dual 
role, 
namely 
the 
ontological role of guiding the particle, and the epistemological 
role of 
giving initially at least the usual prescription for the probability 
density of finding the particle. Also, parameters such as charge, 
mass, 
etc. which are usually associated with the particle are spreadout 
over the wave and not localized on the particle \cite{br95} in the 
Bohm picture. Finally, when one goes over to quantum field 
theory, 
the ontology undergoes a sudden change because the particle is 
replaced by the classical field and it is not clear what its relation is 
to 
the previous ontology.

In the Feynman path integral formalism of quantum theory, the 
measurement problem does not seem to occur, at least not 
explicitly. 
Recently, 
Kaiser and 
Stodolsky \cite{ks95} have claimed that the measurement problem 
does not arise in the Feynman path integral approach. In this 
approach one assumes `events', such 
as the 
spots on a photographic plate, to be a primitive concept. Only 
these 
events are considered to be real. Given an event $A$ caused by a 
system, 
quantum 
mechanics gives the probability ampllitude for a subsequent event 
$B$ to be caused by the same system. This is obtained by 
summing 
the 
probability amplitudes 
associated with the different paths by which the system may go 
from 
$A$ 
to $B$. Here, like in the Copenhagen interpretation, but unlike the 
Everett or Bohm interpretations, the wave function is not real. It is 
the probability amplitude for different possible events, and is 
therefore a prescription for the statistical prediction of these 
events. 
It may then appear that there is no measurement problem 
because 
we can deal directly with probability amplitudes without a wave 
function which undergoes a mysterious collapse.

However, the measurement problem can still be formulated by 
means of the following three questions in the amplitude language, 
with the translation into the wave function language given in 
parentheses. 1) When do we convert probability amplitudes into 
probabilities? (Criterion for macroscopicity of the apparatus?) 2) 
Why 
only one of the many possible events with non zero 
probability amplitude is realized in a particular experiment. 
(Collapse 
problem.) 3) Why don't we see a superposition of the states that 
are 
actually observed in experiments for which also there is a non 
zero 
probability amplitude? (The preferred basis problem.) The wave 
function may be regarded as the probability amplitudes to 
observe 
the particle at various points in space which then relates the 
above 
questions to the corresponding questions in the wave function 
language in parentheses. One cannot make the measurement 
problem 
go away by simply changing the language, which was Wigner's 
answer to my question.

Although I rejected the  Copenhagen interpretation as 
unsatisfactory, 
it may nevertheless be telling us something important. The 
apparatus being `classical' simply means that it should be given a 
space-time description. So, the preferred basis associated with the 
reduction mentioned in the previous section consists of states 
which 
appear `classical', i.e. they have well defined space-time 
representation. The quantum system, on the other hand, has its 
states in the Hilbert space. But the space-time geometry is very 
different from the natural 
geometry for quantum theory which is obtained from the Hilbert 
space (see 
for e.g. \cite{an91}). The 
gravitational field is now incorporated into the geometry of space-
time. Indeed the difficulty in constructing a quantum theory of 
gravity may be due to these very different geometries for space-
time and Hilbert space. But the $R$ process brings these two 
geometries in contact with each other because of the formation of 
events 
when the Hilbert space state vector is observed \cite{an80}. This 
suggests 
that 
the gravitational field may be involved in this process. If the 
gravitational field, which is intimately connected with space-time, 
causes the reduction of the wave packet, then this may explain 
why 
the states into which the collapse takes place have a well defined 
space-time description. Also, this argument suggests that it is not 
necessary to go down to the scale of Planck length for quantum 
gravitational effects to become important, because the above 
problem of relating the Hilbert space geometry to space-time 
geometry, which is required by the reduction of the wave packet, 
exists even 
at much bigger length scales.

\section{Gravitational reduction of the wave packet}

If the wave function is real, as implied by protective observation, 
it 
is likely that its collapse or reduction is also a real process. Also, as 
argued in the previous section, none of the interpretations of 
present 
day quantum theory advanced so far are satisfactory. We should 
therefore be open to the possibility of having to modify quantum 
theory. Several schemes have been proposed without involving 
the 
gravitational field to describe the $R$ process, notably due to 
Pearle~\cite{pea89}, and Ghiradi, Rimini and Weber 
(GRW)~\cite{grw}. 
However, as argued at the end of section 2 it is plausible 
that the gravitational field is involved in the reduction. Indeed 
several suggestions for such a reduction have been made
\cite{ka66,pe81,ggr90}. But I shall consider here only a recent 
specific 
proposal by Penrose~\cite{pe96} which makes the same quantitative 
prediction for the time of collapse as Diosi~\cite{di89}, although 
Penrose's geometrical motivations are different from Diosi's.

To fix our ideas, consider the Stern-Gerlach experiment for a 
spin-half particle such as a neutron. It is well known that as the 
neutron 
passes through the inhomogeneous field of the Stern-Gerlach 
apparatus, its wave function splits into two, and when it interacts 
with a screen the combined wave function of the neutron and the 
screen also splits into two, as they undergo the linear $U$ process 
of quantum mechanics. But the gravitational fields of the two 
states 
are different. So, if the gravitational field is to be treated quantum 
mechanically then the new state is the superposition
\begin{equation}
\Psi = \lambda |\psi_1>|\alpha_1>|\Gamma_1> +\mu 
|\psi_2>|\alpha_2>|\Gamma_2>
=\lambda |\Psi_1>+ \mu |\Psi_2>,
\label{eq:int}
\end{equation}
where $|\psi_1>$ and $|\psi_2>$ are represent the states of the 
neutrons, $|\alpha_1>$ and $|\alpha_2>$ the corresponding 
quantum 
states of the screen with the different positions of the spot where 
the 
neutron strikes, $|\Gamma_1>$ and $|\Gamma_2>$ are the 
coherent states of the gravitational field, and $|\Psi_1>$ and 
$|\Psi_2>$ represent the states of the combined system. 
Interesting 
consequences of a superposition of states of a macroscopic system 
of 
the form (\ref{eq:int}) for a cosmic string have been obtained 
elsewhere \cite{an94,an96}.

Penrose argues \cite{pe96} that in the superposition (\ref{eq:int}) 
there 
must 
necessarily be a `fuzziness' in the time translation operator and a 
corresponding `fuzziness' in the energy. This is important for the 
following reason. In a dynamical collapse model, such as Penrose's 
being considered here, typically there is violation of conservation 
of 
energy-momentum. In the GRW scheme, this violation occurs very 
rarely and so, it was claimed, that it cannot be detected in the 
usual 
experiments. But conservation laws are consequences of 
symmetries, 
which are to me the most fundamental aspects of physics. This is 
illustrated by the fact that although, as mentioned above, the 
Hilbert 
space geometry and space-time geometry are very different, they 
have in common the action of the Poincare symmetry group on 
both 
of them, as if this symmetry group is ontologically prior to both 
descriptions. I expect symmetries of laws of physics and the 
conservation laws which they imply to be more lasting than the 
laws 
themselves. I therefore would not like even a rare violation of the 
conservation of energy-momentum. The `fuzziness' of time 
translation that Penrose mentions, which may be extended also to 
spatial translations, may change the present laws just so as to 
altogether prevent the violation of energy-momentum 
conservation.

The uncertainty of energy associated with this `fuzziness', 
according 
to Penrose, makes superpositions of the form (\ref{eq:int}) 
unstable. 
This is analogous to how the uncertainty of energy $\Delta E$ of a 
particle 
makes it unstable giving it a lifetime is of the order of 
$\hbar\over 
\Delta E$. It is therefore reasonable to suppose that the 
superposed 
states in (\ref{eq:int}) should decay into one or other of the two 
states, 
which we observe to happen in a Stern-Gerlach experiment. The 
lifetime may be postulated to be 
\begin{equation}
T={\hbar\over E}, \label{eq:time}
\end{equation}
where $E$ is to be determined. Penrose considers the special case 
of 
the state $\Psi$ being an equal superposition of two states of a 
lump 
of mass, each of which produces a static gravitational field. In the 
Stern-Gerlach experiment considered above, this corresponds to 
the 
spin state being perpendicular to the inhomogeneous magnetic 
field. 
Define now a quantity which has dimension of energy
\begin{equation}
\Delta = {1\over G}\int (\nabla \Phi_1 - \nabla \Phi_2)^2 d^3x ,
\label{eq:Delta}
\end{equation}
where $\Phi_1$ and $\Phi_2$ are the Newtonian gravitational 
potentials of the two lump states, and $G$ is Newton's 
gravitational 
constant. Penrose \cite{pe96} and Diosi \cite{di89} postulate that $E$ is 
some numerical multiple 
of $\Delta$.

Two questions which arise now are whether this postulate 
can 
be obtained in some natural way and how it could be generalized. 
I shall try to answer both questions. Note first that the 
classical gravitational field corresponds to the mean value of the 
metric operator ${\hat g}_{\mu\nu}$ and the connection operator 
${{\hat\Gamma}_{\mu\nu}}^\rho$. Quantum gravitational effects, 
however, 
depend 
on the fluctuation of the gravitational field. Consider a weak 
gravitational field for which the linearized approximation is 
appropriate. Then the gravitational fields of the superposed states 
may be regarded as perturbations of a background Minkowski 
space-
time. The {\it fluctuation} of the connection $\Delta\Gamma$ is 
given by
\begin{equation}
\Delta\Gamma^2 = \sum\int 
<\Psi|({{\hat\Gamma}_{\mu\nu}}^\rho-
<\Psi|{{\hat\Gamma}_{\mu\nu}}^\rho|\Psi>)^2|\Psi> d^3x  
\label{eq:fluct}
\end{equation}
For (\ref{eq:fluct}) to be physically meaningful, it is necessary to 
eliminate the gauge degrees of freedom by quantizing the 
connection coefficients in an appropriate gauge in which these 
coefficients are unique.
This gauge is here taken to be the gravitational analog of the 
electromagnetic Coulomb gauge that will be defined in the next 
section.
Then the sum in 
(\ref{eq:fluct}) means the summing of the fluctuations of each of 
the operators ${{\hat\Gamma}_{\mu\nu}}^\rho$ defined in this 
gauge to represent the independent physical degrees of freedom.  
Then (\ref{eq:fluct}) may be transformed to any other gravitational 
gauge.

If $\Psi$ is an eigenstate of these operators then (\ref{eq:fluct}) 
vanishes, and the geometry is essentially classical. So, we would 
not 
expect it to decay, i.e. $T$ is infinite.  It is reasonable therefore to 
take $E$ to be proportional to some positive power of 
$\Delta\Gamma$. Since ${1\over G}\Delta\Gamma^2$ has the 
dimension of energy, I postulate that 
\begin{equation}
E = {k\over G}\Delta\Gamma^2 , 
\label{energy}
\end{equation}
where $k$ is some dimensionless constant to be determined by 
the 
future quantum theory of gravity.

Consider now the superposition of two gravitational fields of the 
form (\ref{eq:int}), where $|\Psi_1>$ and $|\Psi_2>$ are 
eigenstates of 
$\hat\Gamma$ with eigenvalues $\Gamma_1$ and $\Gamma_2$. 
Then
\begin{equation}
\Delta\Gamma^2 = \sum\int \{ |\lambda|^2(1-
|\lambda|^2){\Gamma_1}^2 +
|\mu|^2(1-|\mu|^2){\Gamma_2}^2 - 
2|\lambda|^2|\mu|^2\Gamma_1\Gamma_2 \} d^3x.
\label{eq:fluctuation}
\end{equation}
If the fields are Newtonian, then the only non vanishing 
connection coefficients (a la Newton-Cartan theory) are 
$ {{\Gamma_1}_{00}}^i = {\partial 
\Phi_1\over \partial x^i}$ and ${{\Gamma_2}_{00}}^i={\partial 
\Phi_2\over \partial x^i}$ 
Therefore, in the special case considered by Penrose for which 
$\lambda = \mu = {1\over \sqrt 2}$, from (\ref{energy}) and 
(\ref{eq:fluctuation}),
\begin{equation}
E= {k\over 4G}\int (\nabla \Phi_1 - \nabla \Phi_2)^2 d^3x .
\end{equation}
This $E$ is proportional to $\Delta$ given by (\ref{eq:Delta}).

Hence, (\ref{energy}) generalizes Penrose-Diosi ansatz in three 
ways. 
We can now predict the order of magnitude of $T$ for arbitrary 
coefficients $\lambda$ and $\mu$.  Also, (\ref{eq:time}) is valid 
for 
superpositions of more than two lump states. Finally, we can now 
obtain $T$ not only for arbitrary superpositions of static 
gravitational fields but also for non static gravitational fields for 
which there are other components of ${{\Gamma}_{ab}}^c$ besides 
${{\Gamma}_{00}}^i$. For example, the above results may be 
applied to 
Leggett's proposed experiment to realize the quantum 
superposition 
of two currents in a SQUID \cite{le80,ch84,le85}.

The prediction (\ref{eq:time}) together with (\ref{energy}) for the 
time of reduction of the wave packet does not say how this reduction 
takes place. This will be considered in the last section of this paper. 
The question of how well the above 
predictions 
agree with experiment, for example the superposition of two 
currents mentioned above, will be investigated in a future paper. 

\section{Gravitational coulomb gauge}

It has been shown that in electromagnetism the 
suitable gauge for studying the
fluctuation of the vector potential is the 
Coulomb gauge \cite{ah1991}. I shall therefore now define the analog 
of the Coulomb gauge for the gravitational field and require that 
(\ref{eq:fluct}) is defined in this gauge. 

In the weak field limit, we may write $g_{\mu \nu}=\eta_{\mu 
\nu}+\gamma_{\mu \nu}$, where $\gamma_{\mu \nu}<<1$. Then I 
shall define the gravitational Coulomb gauge by 
\begin{equation}
\sum_{j=1}^3{\gamma^{\mu j}}_{,j}=0
\label{coulomb}
\end{equation}
The linearized Einstein field equations are \cite{mi1973}
\begin{equation}
{{\gamma}_{\mu\alpha,\nu}}^\alpha+
{{\gamma}_{\nu\alpha,\mu}}^\alpha 
-{{\gamma}_{\mu\nu,\alpha}}^\alpha -{\gamma}_{,\mu\nu}-
{\eta}_{\mu\nu}({{\gamma}_{\alpha\beta,}}^{\alpha\beta}-
{{\gamma}_{,\beta}}^\beta)=16\pi T_{\mu\nu}
\label{linear}
\end{equation}
where $,\mu$ represents partial derivative with respect to $x^\mu$, 
$\gamma={\gamma_\alpha}^\alpha$ and repeated greek indices are 
summed over $0,1,2,3$. The metric has signature $(-+++)$.
On imposing the gauge condition 
(\ref{coulomb}), (\ref{linear}) reads
\begin{equation}
{{\gamma}_{\mu 0,\nu}}^0+
{{\gamma}_{\nu 0,\mu}}^0 -{{\gamma}_{\mu\nu,\alpha}}^\alpha 
-{\gamma}_{,\mu\nu}-
{\eta}_{\mu\nu}({\gamma}_{00,00}-
{{\gamma}_{,\beta}}^\beta)=16\pi T_{\mu\nu}
\label{linear1}
\end{equation}
The $(\mu,\nu)=(0,0), (i,0)$ and $(i,j)$ components of (\ref{linear1}) 
are respectively
\begin{equation}
-\sum_{k=1}^3 \sum_{m=1}^3 {\gamma}_{kk,mm} =16\pi T_{00} ,
\label{linear2}
\end{equation}
\begin{equation}
-\sum_{j=1}^3({\gamma}_{i0,jj} + {\gamma}_{jj,i0}) =16\pi
T_{i0}
\label{linear3}
\end{equation}
and
\begin{equation}
{{\gamma}_{i 0,j}}^0+
{{\gamma}_{j0,i}}^0 -{{\gamma}_{ij,\alpha}}^\alpha +{\gamma}_{00,ij}-
\sum_{k=1}^3 h_{kk,ij}-{\delta}_{ij}\sum_{k=1}^3 
({\gamma}_{00,kk}-{{\gamma}_{kk,\beta}}^\beta)=16\pi T_{ij}
\label{linear4}
\end{equation}

I assume now that each $\gamma_{\mu\nu}$ falls off sufficiently 
rapidly at infinity so that the Poisson's 
(\ref{linear2}) determines $\sum_{k=1}^3 
{{\gamma}_{kk}}$ in terms of $T_{00}$ uniquely. Substituting this 
into (\ref{linear3}), $\gamma_{i0}$ is determined uniquely in terms 
of $T_{00}$ and $T_{i0}$. Now take trace of (\ref{linear4}) by 
summing over $i=j=1,2,3$. Substitute for $\sum_{k=1}^3 
{{\gamma}_{kk}}$  and $\gamma_{i0}$ the values we have just 
found. Then $\gamma_{00}$ satisfies Poisson equation with the 
source being a function of $T_{\mu\nu}$. Hence, $\gamma_{00}$ is 
determined uniquely. 

It is easy to show that the remaining components 
$\gamma_{ij}~(i,j=1,2,3)$ of 
$\gamma_{\mu\nu}$ are determined uniquely by (\ref{coulomb}).
To see this, do an infinitesimal gauge transformation 
$x^{\mu\prime} = x^\mu -\xi^\mu$. 
Then the corresponding transformation of $g_{\mu\nu}$ 
is equivalent to the transformation 
\begin{equation}
\gamma_{\mu\nu}'=\gamma_{\mu\nu}+\xi_{\mu,\nu}+
\xi_{\nu,\mu}.
\label{diffeo}
\end{equation} 
Considering $(\mu,\nu)=(i,j)$ and requiring (\ref{coulomb}) in the 
new gauge also,
\begin{equation}
\sum_{j=1}^3(\xi_{i,jj}+\xi_{j,ij})=0.
\label{gauge1}
\end{equation}
In momentum space\footnote{This argument was made in 
collaboration with Joseph Samuel.} (\ref{gauge1}) reads 
\begin{equation}
{\bf k}^2 \tilde{\xi}^i +({\bf k}\cdot{\bf \tilde\xi})k^i =0,
\label{gauge2}
\end{equation}
where $\tilde{\xi}^i$ is the Fourier transform of $\xi^i$.
Therefore, $\tilde{\xi}^i $ is proportional to $k^i$ and so we can write 
$\tilde{\xi}^i =\alpha({\bf k}) k^i$. Substituting into (\ref{gauge2}),
either $\alpha=0$ or ${\bf k}^2=0$, which means that $\xi^i$ has no 
spatial dependence. It follows that, in either case, from (\ref{diffeo}), 
$\gamma_{ij}'=\gamma_{ij}$. But it was shown earlier that
$\gamma_{\mu 0},\mu=0,1,2,3$ are uniquely determined in this gauge. Hence,
all $\gamma_{\mu\nu}$ are uniquely determined in the gauge
(\ref{coulomb}).

An interesting aspect of this gauge is that in the absence of matter, 
the above results imply
\begin{equation}
\gamma_{\mu 0}=0, \mu=0,1,2,3, ~\sum_{k=1}^3 {{\gamma}_{kk}}=0, 
\sum_{j=1}^3\gamma_{i j,j}=0.
\end{equation}
I.e. the above gravitational coulomb gauge
reduces to the usual transverse traceless gauge in the absence of 
matter, which is the physical gauge for gravitational radiation.

\section{Laws, symmetry and the measurement process}

It was mentioned at the beginning of section 2 that quantum 
theory does not explain the occurrence of events. `Explain' here 
impllicitly assumes having a causal law that describes the formation 
of events. But the very notion of law is strange in that it carries with 
it a necessity which is not logical or mathematical necessity. This is 
because a law must be refutable, whereas a logical or mathematical 
necessity is tautological and therefore cannot be refuted. It is 
therefore reasonable to consider the consequences of there not being 
any laws of necessity.

A 
law of necessity, or simply a law, may be defined as the ability to 
describe the initial 
state of a physical system in such a way that the final state may be 
predicted uniquely
from this initial state using the nature of the system and its 
interaction 
with its environment. 
The absence of laws then implies that identical 
physical systems may start from the same initial state and end up in 
different final states. This statement is consistent with our 
observation of quantum phenomena. However, as mentioned in 
section 3, attempts were made to violate this statement and 
make quantum theory conform to the paradigm that 
all phenomena occur according to laws.

By means of protective measurements \cite{aav93}, which we can do 
in principle, any state in the Hilbert space can be observed for a 
single system. I shall therefore allow any state in the Hilbert space  
to be the initial or final state of a system. However, we observe 
macroscopic systems in states in which the wave packets of its 
constituents are localized. Hence, even if it is initially in a state
whose 
wave function is spread out then it could end up in a state which is 
sufficiently well localized. According to the hypothesis advanced 
here, this process is not described by a deterministic law. 
Nevertheless, the prediction (\ref{eq:time}) together with 
(\ref{energy}) will apply to the time taken for the initial state to 
become the final state.

But in order to give up laws, it is necessary to provide an alternative 
explanation for the regularities in the phenomena which we observe. 
E.g. why do planets have seemingly precise orbits? 
Or why are there precise experimentally well confirmed 
probabilities for the possible final states of a given 
initial state of a quantum system? I 
believe that regularities such as these may be explained by 
symmetries. 

As for the first question, the seemingly  precise motions in 
classical physics must be obtained as appropriate limiting cases 
of the quantum motion, which so far has been described by 
the motion of a wave function. The external field modifies 
this motion by means of phase shifts in interfering secondary 
wavelets in the propagation of this wave via Huygen¹s principle.
But it was shown that  the phase shifts due to gravity and gauge 
fields are caused by elements of the Poincare group and the 
corresponding gauge group, respectively \cite{an1979,an96}. 
These phase shifts of course were obtained from laws which 
have these symmetries. But once having obtained them we could 
give up the laws and keep only the symmetries.

As for the second question, consider the tossing of a coin. The 
equal probabilities for `heads' 
and `tails' is due to the symmetry between them. The probabilities 
are independent of the particular law governing the motion of the 
coin, so long as this law respects this symmetry. This suggests that 
the probabilities may be independent of the existence of laws, and 
governed only by the symmetries. It may be possible to deduce the 
quantum mechanical probabilities by symmetry considerations alone.

If we give up the laws then the problem of energy-momentum 
non conservation in the dynamical collapse models, which 
was mentioned in section 3, disappears. We simply accept the 
transition of a quantum state from a given initial state to a final state, 
without a law governing this process, in such a way that 
energy-momentum and all other conserved quantities 
corresponding to the 
known symmetries are conserved for the combined system 
consisting of each observed system and its environment. 
In quantum theory there is a direct connection between the 
symmetries and conserved quantities due to the symmetries 
that act on the Hilbert space being generated by the conserved
quantitites.

Another problem with dynamical collapse, if we work within the
paradigm of laws, is that if a charged particle wave function 
collapses we   would expect it to radiate, which has not been observed. 
But if we
give up laws and work in a new paradigm in 
which symmetries are the fundamental concepts then again we can 
simply accept the transition of this wave function from the initial 
to the final state without requiring a law to account for the motion 
in between. This transition then could happen without any radiation.

But a great deal of work remains to show that all the regularities that 
we observe in nature could be obtained from symmetries and logical 
or mathematical self consistency. This will be explored in 
another paper \cite{an1998}.

\vskip .5cm
\noindent
Acknowledgments
\vskip .5cm
I thank Yakir Aharonov, Abhay Ashtekar and Roger Penrose for 
useful discussions. 
This work 
was supported by ONR grant no. R\&T 3124141 and NSF grant no. 
PHY-9307708.

\end{document}